\documentclass[prx,a4paper,twocolumn,showpacs,floatfix,superscriptaddress,amsmath,amsfonts,amssymb,preprintnumbers,citeautoscript,aps,longbibliography]{revtex4}
\pdfoutput=1

\usepackage[T1]{fontenc}\usepackage[latin1]{inputenc}
\usepackage{dcolumn,graphicx,color,booktabs,microtype,afterpage}
\graphicspath{{Figures/}}
\usepackage[charter,greekuppercase=italicized]{mathdesign}
\usepackage{sidecap}
\usepackage{bm}
\usepackage{braket}
\renewcommand{\figurename}{Fig.}
\renewcommand{\tablename}{Table}
\makeatletter\renewcommand{\fnum@figure}[1]{\figurename~\thefigure.}\makeatother
\makeatletter\renewcommand{\fnum@table}[1]{\tablename~\thetable.}\makeatother

\newcount\hh \newcount\mm
\hh=\time \divide\hh by 60
\mm=\hh \multiply\mm by 60 \mm=-\mm
\advance\mm by \time
\def\now{\number\hh:\ifnum\mm<10{}0\fi\number\mm}

\usepackage[colorlinks,plainpages=false,linkcolor=blue,urlcolor=blue,citecolor=blue,pdfpagemode=UseNone,pdfstartview=FitBH]{hyperref}

\newcommand{\McSymbol}{\includegraphics[height=6.8pt]{McSymbol.pdf}\kern0.5pt}

\begin{document}

\makeatletter\renewcommand{\ps@plain}{%
\def\@evenhead{\hfill\itshape\rightmark}%
\def\@oddhead{\itshape\leftmark\hfill}%
\renewcommand{\@evenfoot}{\hfill\small{--~\thepage~--}\hfill}%
\renewcommand{\@oddfoot}{\hfill\small{--~\thepage~--}\hfill}%
}\makeatother\pagestyle{plain}


\title{Magnonic Weyl states in Cu$_2$OSeO$_3$}

\author{L.~Zhang}
\affiliation{Peter Gr\"unberg Institut and Institute for Advanced Simulation, Forschungszentrum J\"ulich and JARA, 52425 J\"ulich, Germany}
\affiliation{Department of Physics, RWTH Aachen University, 52056 Aachen, Germany}

\author{Y.~A. Onykiienko}
\affiliation{Institut f\"ur Festk\"orper- und Materialphysik, Technische Universit\"at Dresden, 01069 Dresden, Germany}

\author{P.~M.~Buhl}
\affiliation{Peter Gr\"unberg Institut and Institute for Advanced Simulation, Forschungszentrum J\"ulich and JARA, 52425 J\"ulich, Germany}
\affiliation{Department of Physics, RWTH Aachen University, 52056 Aachen, Germany}

\author{Y.~V. Tymoshenko}
\affiliation{Institut f\"ur Festk\"orper- und Materialphysik, Technische Universit\"at Dresden, 01069 Dresden, Germany}

\author{P. \v{C}erm\'{a}k}
\affiliation{Forschungszentrum J\"ulich GmbH, J\"ulich Center for Neutron Science at MLZ, Lichtenbergstr.~1, 85748 Garching, Germany}
\affiliation{Charles University, Faculty of Mathematics and Physics, Department of Condensed Matter Physics, Ke Karlovu 5, 121~16, Praha, Czech Republic}

\author{A.~Schneidewind}
\affiliation{Forschungszentrum J\"ulich GmbH, J\"ulich Center for Neutron Science at MLZ, Lichtenbergstr.~1, 85748 Garching, Germany}

\author{A.~Henschel}
\affiliation{Max Planck Institute for Chemical Physics of Solids, N\"othnitzer Stra{\ss}e 40, 01187 Dresden, Germany}

\author{M.~Schmidt}
\affiliation{Max Planck Institute for Chemical Physics of Solids, N\"othnitzer Stra{\ss}e 40, 01187 Dresden, Germany}

\author{S.~Bl\"ugel}
\affiliation{Peter Gr\"unberg Institut and Institute for Advanced Simulation, Forschungszentrum J\"ulich and JARA, 52425 J\"ulich, Germany}

\author{D.~S.~Inosov}\email[Corresponding author:~]{dmytro.inosov@tu-dresden.de}
\affiliation{Institut f\"ur Festk\"orper- und Materialphysik, Technische Universit\"at Dresden, 01069 Dresden, Germany}

\author{Y.~Mokrousov}\email[Corresponding author:~]{y.mokrousov@fz-juelich.de}
\affiliation{Peter Gr\"unberg Institut and Institute for Advanced Simulation, Forschungszentrum J\"ulich and JARA, 52425 J\"ulich, Germany}
\affiliation{Institute of Physics, Johannes Gutenberg University Mainz, 55099 Mainz, Germany}

\begin{abstract}
\noindent The multiferroic ferrimagnet Cu$_2$OSeO$_3$ with a chiral crystal structure attracted a lot of recent attention due to the emergence of magnetic skyrmion order in this material. Here, the topological properties of its magnon excitations are systematically investigated by linear spin-wave theory and inelastic neutron scattering. When considering Heisenberg exchange interactions only, two degenerate Weyl magnon nodes with topological charges $\pm 2$ are observed at high-symmetry points. Each Weyl point splits into two as the symmetry of the system is further reduced by including into consideration the nearest-neighbor Dzyaloshinsky-Moriya interaction, crucial for obtaining an accurate fit to the experimental spin-wave spectrum. The predicted topological properties are verified by surface state and Chern number analysis. Additionally, we predict that a measurable thermal Hall conductivity can be associated with the emergence of the Weyl points, the position of which can be tuned by changing the crystal symmetry of the material.
\end{abstract}

\maketitle

\section{\hspace{-1ex}Introduction}\label{Sec:Introduction}

Topological insulators and Weyl semimetals attracted tremendous attention as the most prominent realizations of topologically nontrivial electronic matter \cite{Bernevig13, OrtmannRoche15, BansilLin16, ArmitageMele18, ZangCros18}. In recent years, topologically protected band touching points, known as Weyl nodes, were observed in electronic~\cite{XuBelopolski15, LvWeng15, LuWang15}, photonic~\cite{LuJoannopoulos14}, phononic~\cite{LiHuang18}, and magnetic excitation spectra~\cite{ChisnellHelton15, YaoLi18, BaoWang18}. In relationship to magnetically ordered materials, new concepts of topological magnon insulators \cite{ZhangRen13, Owerre16, NakataKim17, LiKovalev18}, topological spinon semimetals \cite{SchafferLee15}, Dirac and Weyl magnon states \cite{LiLi16, MookHenk16, LiLi17, JianNie18, Owerre18} were introduced, offering promising new applications in the emerging field of spintronics~\cite{SmejkalJungwirth17, SmejkalMokrousov18, RuckriegelBrataas18, WangZhang18}.

Experimentally, topologically nontrivial magnon states were recently identified in a two-dimensional \mbox{spin-$\frac{1}{2}$} kagome-lattice ferromagnet \cite{ChisnellHelton15} and in the three-dimensional (3D) antiferromagnet Cu$_3$TeO$_6$ by two independent groups~\cite{YaoLi18, BaoWang18} using inelastic neutron scattering (INS). On the theory side, it has been realized that chiral magnets offer a generic route to the realization of topological magnon states, representing a magnon analog of topological insulators. As a result of antisymmetric exchange, known as Dzyaloshinsky-Moriya interaction (DMI)~\cite{Dzyaloshinsky58, Moriya60}, the bulk magnon spectrum of a chiral magnet can acquire a topological energy gap that supports a topologically protected gapless Dirac cone in the surface magnon spectrum \cite{LiKovalev18}. A similar mechanism based on DMI was also proposed for the formation of magnonic Weyl crossing points in the spin-wave spectrum of the noncoplanar antiferromagnetic (AFM) state on a breathing-pyrochlore lattice \cite{LiLi16, MookHenk16, JianNie18}.

The cubic copper(II)-oxoselenite Cu$_2$OSeO$_3$ is a multiferroic ferrimagnet with a chiral crystal structure that came under the focus of recent attention owing to the emergence of skyrmion order in this material~\cite{SekiYu12, SekiKim12, LangnerRoy14, LangnerRoy17, MuellerRajeswari17}. Its crystal structure is cubic (space group $P2_{1}3$) with the lattice constant $a=8.925$\,\AA~\cite{EffenbergerPertlik86}. The magnetic sublattice of Cu$^{2+}$ ions can be approximated as a distorted breathing-pyrochlore lattice, consisting of slightly deformed tetrahedral Cu$_4$ clusters in a face-centered cubic (fcc) arrangement \cite{PortnichenkoRomhanyi16}. Magnetic interactions within the tetrahedron lead to a ferri\-magnetic ground state, in which one of the Cu$^{2+}$ spins is antiparallel to the other three, resulting in the total spin $S=1$ of the cluster~\cite{YangLi12, Romhanyi14, Janson14}. Weaker interactions between the clusters lead to a long-range spin-spiral order that sets in below $T_\text{C}\approx57$\,K. Existing magnetic models \cite{YangLi12, Romhanyi14, Janson14, ChizhikovDmitrienko15} consider up to 5 Heisenberg exchange interactions and up to 5 DMI vectors. These models were used to describe the INS spectrum of spin-wave excitations in a broad energy range and in the whole reciprocal space \cite{PortnichenkoRomhanyi16}, as well as electron spin resonance (ESR) that probes spin-wave excitations at the zone center \cite{OzerovRomhanyi14}. However, the DMI was initially neglected in these studies.

This simplified description, that involves only Heisenberg interactions, provides a qualitatively good fit to the experimental spin-wave dispersion over the entire Brillouin zone~\cite{PortnichenkoRomhanyi16} with the exception of the zone corner ($\mathbf{R}$ point), where the magnon bands remain degenerate for any values of the exchange parameters. Tucker \textit{et al.} \cite{TuckerWhite16} recently showed that this degeneracy is removed by DMI, leading to a clearly resolved spin gap of $\sim$1.6~meV in the magnon spectrum, which they observed by neutron spectroscopy. These observations are a strong indication for the existence of topological magnon states in Cu$_2$OSeO$_3$, which motivated our present study.

In the following, we present spin-dynamical calculations of the magnon spectrum in the presence of DMI that was adjusted to provide the best fit to the experimental spin-wave dispersion in the vicinity of the $\mathbf{R}$ point. Using linear spin-wave theory (LSWT) in combination with high-resolution neutron spectroscopy, we show that, in the absence of DMI terms, two pairs of degenerate Weyl nodes with the topological charge $+2$ and $-2$ are located at the zone center ($\mathbf{\Gamma}$ point) and at the zone boundary ($\mathbf{R}$ point). Consideration of the nearest-neighbor DMI is sufficient to lift the degeneracy of these Weyl nodes, so that they are shifted away from the high-symmetry points into a position that sensitively depends on the direction and magnitude of the DMI vector. A direct observation of the resulting Weyl points would offer a possibility to accurately extract the DMI from INS measurements. We verify the predicted topological properties by the Chern number analysis and give quantitative predictions for the location of magnonic Weyl points in the spin-wave spectrum. We also analyze topologically protected magnon surface states and estimate the magnetic contribution to the thermal Hall conductivity that may serve as robust hallmarks of the emergent topological states in Cu$_2$OSeO$_3$, awaiting a direct experimental verification.\vspace*{-5pt}

\section{\hspace{-1ex}Results}\label{Sec:Results}\vspace*{-3pt}
\subsection{\hspace{-1ex}Magnetic model, experimental result and magnon spectrum}\vspace{-2pt}

\begin{figure*}[t!]
\includegraphics[width=\textwidth]{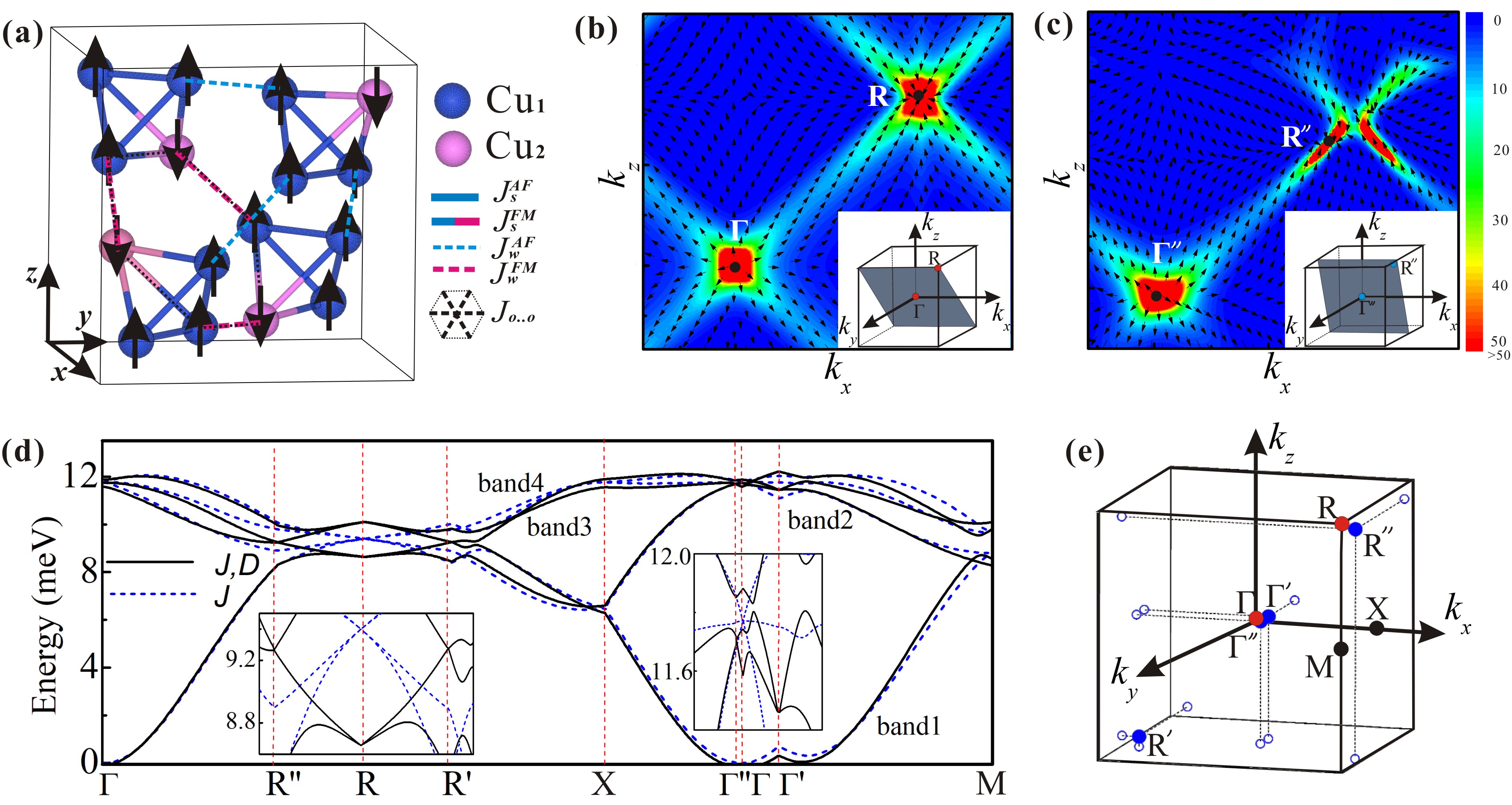}
\caption{(a) The structure of ferrimagnetic Cu$_2$OSeO$_3$ is shown together with five different interactions paths,  marked by $J_\text{s}^\text{FM}$,  $J_\text{s}^\text{AFM}$, $J_\text{w}^\text{FM}$, $J_\text{w}^\text{AFM}$ and $J_\text{O..O}$, where $J_\text{O..O}$ represents the antiferromagnetic long-range interaction. (b\,--\,c) Monopole distribution of the absolute magnitude of the Berry curvature corresponding to two lowest magnon bands in the $k$-planes marked in the inset. In (b), the DMI was not taken into account, while (c) shows the result with the DMI included. The $k$-planes are chosen so that they include the Weyl points. (d) The magnon dispersion of the lowest four bands of Cu$_2$OSeO$_3$. The dotted blue line represents the dispersion without the effect of the DMI, with the Weyl points emerging at $\mathbf{R}$ and at $\mathbf{\Gamma}$. The black line represents the dispersion upon including the DMI, with four Weyl points emerging at $\mathbf{R^\prime}$, $\mathbf{R^{\prime\prime}}$, $\mathbf{\Gamma^\prime}$ and $\mathbf{\Gamma^{\prime\prime}}$. The exact positions of the Weyl points are shown in (e): $\mathbf{R^{\prime\prime}}=(0.46, 0.19, 0.42)$, $\mathbf{R^{\prime}}=(0.57, 0.18, 0.54)$, $\mathbf{\Gamma^{\prime\prime}}=(0.01, -0.02, -0.01)$, $\mathbf{\Gamma^{\prime}}=(0.01, -0,13, -0.01)$, their unit-direction projections are indicated by blue open circles.
}
    \label{fig:1}
\end{figure*}
The crystal structure of Cu$_2$OSeO$_3$ belongs to the chiral space group $P2_13$ and contains 16 magnetic Cu$^{2+}$ ions per unit cell with $S=\frac{1}{2}$. They occupy two structurally nonequivalent positions, so that every Cu$_4$ tetrahedral cluster consists of one Cu(1) ion on the 4$a$ Wyckoff site and three Cu(2) ions on the 12$b$ site \cite{EffenbergerPertlik86, BosColin08}, see Fig.~\ref{fig:1}. The strong superexchange coupling $J_\text{s}^\text{FM}$ between the Cu(2) ions within the cluster is ferromagnetic (FM), whereas the Cu(1) and Cu(2) spins within the same tetrahedron are coupled antiferromagnetically with a coupling constant $J_\text{s}^\text{AFM}$. These exchange constants constitute the dominant magnetic interactions that lead to a ferrimagnetic spin arrangement within the cluster: Three Cu(2) spins align ferromagnetically, and the Cu(1) spin is pointing in the opposite direction, resulting in a total spin of $S$\,=\,1~\cite{BelesiRousochatzakis10}. The intercluster interactions are considerably weaker, given by the FM superexchange $J_\text{w}^\text{FM}$ between the nearest Cu(2) ions of neighboring clusters, the weak AFM coupling $J_\text{w}^\text{AFM}$ between Cu(1) and Cu(2), and a longer-range exchange $J_\text{O..O}^\text{AFM}$ that connects Cu(1) and Cu(2) sites across the diagonals of alternating Cu(1)--Cu(2) hexagon loops~\cite{Romhanyi14}, see Fig.~\ref{fig:1}.

Numerical values of all five Heisenberg interactions have been calculated from the microscopic electronic structure theory and verified using thermodynamic data \cite{Janson14}, tera\-hertz ESR \cite{OzerovRomhanyi14}, far-infrared \cite{ChizhikovDmitrienko15} and Raman \cite{MillerXu10} spectroscopy, and INS measurements \cite{PortnichenkoRomhanyi16, TuckerWhite16} in earlier works. As a result, there are accurate quantitative estimates of all five exchange parameters. On the other hand, antisymmetric DMI is also allowed by crystal symmetry along all mentioned exchange paths.
Each DMI channel can add at most three extra parameters,
which are the off-diagonal components of the $\hat{J}$ tensor. This results in up to 15 additional parameters in the magnetic Hamiltonian.
All previously reported attempts to estimate their magnitude  are based on first principles calculations \cite{Janson14}.
Hence, the measurable DMI-signatures in the spin-wave spectrum \cite{TuckerWhite16} still await experimental verification.

Here we use LSWT to calculate the magnon spectrum of Cu$_2$OSeO$_3$, starting from the generalized Heisenberg model,
\begin{equation}
  H=\sum_{\langle ij\rangle}\mathbf{S}_i^{\dagger}\hat{J}_{ij}\mathbf{S}^{\phantom{\dagger}}_j
  \label{equ1}\vspace{-1ex},
\end{equation}
where the interaction tensor between the lattice sites $i$ and $j$
\begin{equation}
\hat{J}_{ij}={
\left(
 \begin{array}{ccc}
 J_{ij}^{x}  &  D_{ij}^{z} &  -D_{ij}^{y}\\
 -D_{ij}^{z} &  J_{ij}^{y}  & D_{ij}^{x}\\
 D_{ij}^{y}  & -D_{ij}^{x}  & J_{ij}^{z}
 \end{array}
  \right)}
\end{equation}
includes the symmetric exchange $\mathbf{J}_{ij}$ and the antisymmetric off-diagonal DMI terms $\mathbf{D}_{ij}$, caused by the spin-orbit coupling.
The DMI vector is defined as $\mathbf{D}_{ij}=(D_{ij}^x, D_{ij}^y, D_{ij}^z)$. Following earlier works \cite{PortnichenkoRomhanyi16}, we include five Heisenberg exchange interactions shown in Fig.\,\ref{fig:1}(a), with their numerical values listed in Table~\ref{table:1}. To deal with the ferrimagnetic system, the rotation matrix $O_i$ is introduced, where $O_i$ determines the magnetic moment direction at the site $i$.

To get the magnons excitation spectrum, the LSWT is used\,\cite{TothLake15,LiLi17,SantosSantosDias18,owerre2018weyl}, where the Holstein-Primakoff transformation~\cite{holstein1940field} is adopted for the quantum spin operators. A Fourier transformation of the boson operators is given by
\begin{equation}
   \alpha_l(\mathbf{k})=
\left(
 \begin{array}{cc}
  a_{l}(\mathbf{k})  \\
  a_{l}^{\dagger}(\mathbf{-k})
 \end{array}
  \right)
  = \frac{1}{\sqrt{N}}\sum_m e^{-\mathrm{i}\mathbf{k}\mathbf{R_m}}
  \left(
 \begin{array}{cc}
  a_{lm}  \\
  a_{lm}^{\dagger}
 \end{array}
  \right),
\end{equation}
 where $N$ is the number of the unit cells, and $\mathbf{k}$ is the vector in the reciprocal $k$-space of magnons (suppressed below for convenience).
 The fourier-transformed Hamiltonian part quadratic in the Boson operators, denoted as $H_2$, becomes a $2n\times 2n$ matrix, where $n$ is the number of the atoms in the unit cell.
 From the commutation relation between the bosonic creation (annihilation) operators and $H_2$, we arrive at the equation
\begin{equation}
    \mathrm{i}\frac{\mathrm{d}\alpha_l}{\mathrm{d}t}=[\alpha_l,H_2]=\sum_mD_{lm}\alpha_m,
\end{equation}
where the dynamical matrix is given by $D=gH_2$ with $g=[(\mathbb{1}, 0), (0, -\mathbb{1})]$,
where $\mathbb{1}$ is the $n \times n$ identity matrix.
The positive real eigenvalues of the  dynamical matrix $D$ correspond to the magnons excitation spectrum in the system.
While the left and right eigenvectors of $D$, denoted as $V_L$ and $V_R$, may differ since $D$ is not necessarily hermitian, their relation is trivially given by $V_L= gV_R^{\dagger} g$.

The INS experiment has been carried out on the cold-neutron triple-axis spectrometer PANDA \cite{SchneidewindCermak15} located at MLZ in Garching, Germany. The sample is a coaligned mosaic of 11 single crystals with a total mass of $\sim$\,2 grams and a mosaicity of $\sim$\,2$^\circ$. It was mounted in the $(HHL)$ scattering plane, i.e. with the $[1\overline{1}0]$ axis vertical, inside the JVM1-5.0T cryomagnet with a base temperature of 1.5~K. The instrument was operated with a fixed final neutron wavelength $k_\text{f} = 1.5$~\AA$^{-1}$. To avoid higher-order contamination from the monochromator, a cold beryllium filter was mounted between the sample and the analyzer.

\begin{figure*}
    \centering
    \includegraphics[width=\linewidth]{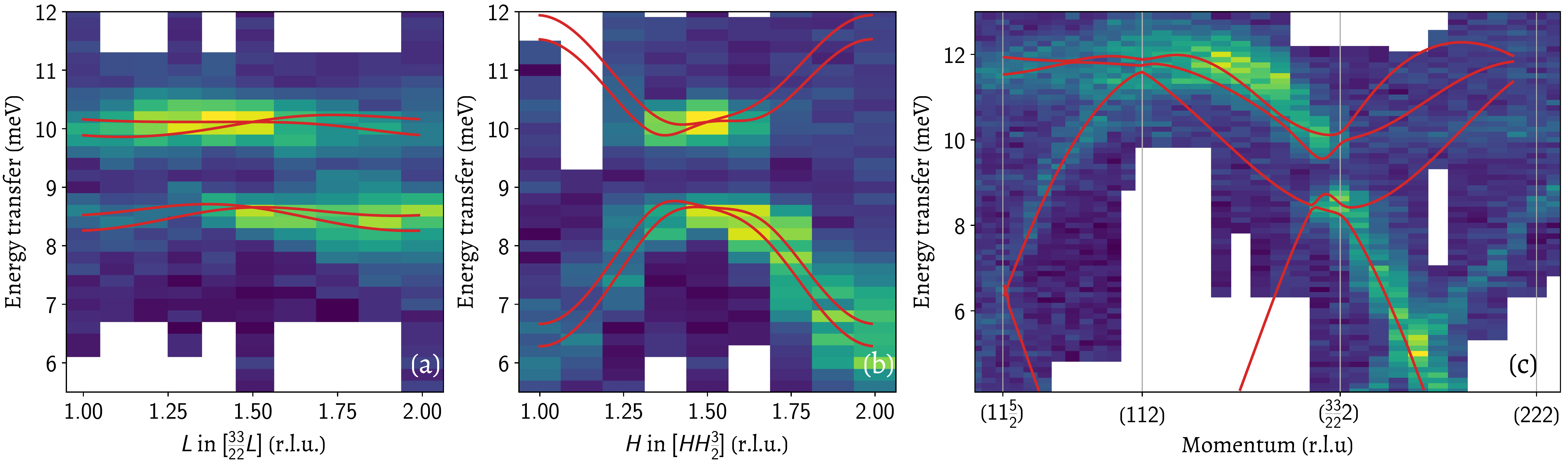}
    \caption{Momentum-energy cuts along (a) $(HH\frac{3}{2})$, (b) $(\frac{3}{2}\frac{3}{2}L)$, and (c) $(11L)$--$(HH2)$ directions in reciprocal space. The first two cuts are centered a the $\mathbf{R}(\frac{3}{2}\frac{3}{2}\frac{3}{2})$ point, where degenerate magnon bands are predicted in the absence of DMI. All measurements were done at 2~K without applying magnetic field. The magnon dispersion calculated with fitted value of the DMI is shown with thin red lines. The straight feature in panel (c), which is not captured by the spin-wave model, is a spurious feature from nonmagnetic multiple scattering.}
    \label{fig:2}
\end{figure*}

The magnon spectrum of Cu$_2$OSeO$_3$ has been analyzed in several previous works, which used similar values of the exchange parameters $J$ and neglected the effect of the DMI~\cite{Romhanyi14, TuckerWhite16, PortnichenkoRomhanyi16}. In these works, similar magnon dispersions were obtained, featuring two doubly-degenerate crossing points: one at $\mathbf{\Gamma}$ and one at $\mathbf{R}$ high-symmetry points. The results of our calculations, performed without DMI, are very close to previously published data, and they are shown in Fig.~\ref{fig:1}\,(d).
\begin{table}[b!]
\begin{tabular}{l@{~~}l@{~~~}r@{~\hspace{2em}~}r}
    \toprule
        Parameters & Distance~(\AA)  & $J$\,(meV)~\cite{PortnichenkoRomhanyi16}\hspace{-2em} & $\mathbf{D}$\,(meV) \\
	\midrule
    	$J_\text{w}^\text{FM}$ & 3.039 & --4.2~~ & (--0.458,\,2.011,\,0.565)\smallskip\\
        $J_\text{s}^\text{AFM}$ & 3.057 & 12.3~~ & 0\smallskip\\
        $J_\text{s}^\text{FM}$ & 3.22 & --14.5~~ & 0\smallskip\\
        $J_\text{w}^\text{AFM}$ & 3.30 & 2.33 & 0\smallskip\\
        $J_\text{O..O}^\text{AFM}$ & 6.35 & 3.88 & 0\\
    \bottomrule
\end{tabular}
\caption{Values of the parameters entering the Heisenberg Hamiltonian~(1). Five exchange interactions are listed together with the corresponding interatomic distances which are very similar to previous works. The nearest-neighbor DMI vector was chosen so as to reproduce the experimental spin-wave dispersion.\vspace{-3pt}}
\label{table:1}
\end{table}
On the other hand, the results of our high-resolution INS measurements, presented in Fig.\,\ref{fig:2}, clearly mark the formation of a 1.6~meV band gap at $\mathbf{R}$ between the bands 2 and 3, according to the enumeration of Fig.~\ref{fig:1}\,(d).
To reproduce this band gap in the calculations, we have chosen the value of the nearest-neighbor DMI by fitting it to the experimental data.
The values of fitted $J$ and $D$ parameters are listed in Table\,\ref{table:1}, whereas Fig.\,\ref{fig:2} shows the comparison of the experimental and calculated magnon spectra. The value of the DMI that we use in this work provided by far the best fit to the experiment among other possible DMI choices, which e.g. included more neighbors into considerations, or even with respect to  previously published {\it ab-initio} results for the DMI in this system~\cite{Janson14}, see Figs.~S2 and S3 of the Supplemental Material \cite{SupplementalMaterial}.
Irrespective of its exact choice, including the DMI into the picture has a drastic effect on the number and position of the degenerate crossings between bands 2 and 3 in the magnonic band structure.
The set of DMI parameters we used here (see Table~\,\ref{table:1}), splits previously degenerate crossing points at $\mathbf{\Gamma}$ and $\mathbf{R}$, giving rise to overall four crossings:
two at $\mathbf{R^{\prime}}$ and $\mathbf{R^{\prime\prime}}$ (in the vicinity of $\mathbf{R}$), and two at $\mathbf{\Gamma^{\prime}}$ and $\mathbf{\Gamma^{\prime\prime}}$ (in the vicinity of $\mathbf{\Gamma}$), see Fig.\,\ref{fig:1}\,(d,e). In the next subsection we analyze the topological character of these points.

\subsection{\hspace{-1ex}Topological properties}\vspace{-2pt}
In order to access the topological properties of the system we calculate the Berry curvature of each magnonic band $n$, defined as \cite{MookHenk16, owerre2018weyl}
\begin{equation}
 \boldsymbol{\mathrm{\Omega}}_n(\mathbf{k})=-\sum_{m\neq n}
 \frac{\mathrm{Im}\left[\braket{V^{L}_{n\mathbf{k}}|\partial_{\mathbf{k}} D(\mathbf{k})|V^{R}_{m\mathbf{k}}}\times\braket{V^{L}_{m\mathbf{k}}|\partial_{\mathbf{k}}D(\mathbf{k})|V^{R}_{n\mathbf{k}}}\right]}{(\epsilon_{n\mathbf{k}}-\epsilon_{m\mathbf{k}})^2},
\end{equation}
where $\epsilon_{m\mathbf{k}}$ are the magnonic eigenvalues.
As we mainly focus on the topological nature of the band crossings arising between the bands $2$ and $3$, we analyze the cumulative Berry curvature of the bands $1$ and $2$. In Fig.\,\ref{fig:1}\,(b) we present the direction of the normalized projected cumulative Berry curvature vector field and its absolute magnitude in the $k_y=k_z$ plane, first for the case without DMI. In the mentioned figure, the color scale represents the absolute value of the Berry curvature vector field. As apparent from the figure, the Berry curvature distribution exhibits two monopole-like features at $\mathbf{R}$ and $\mathbf{\Gamma}$, where the band crossings occur, with the crossing at $\mathbf{\Gamma}$ serving as a source, and the crossing at $\mathbf{R}$
serving as a sink of the Berry curvature field. The corresponding distribution, obtained after including the DMI, is shown in Fig.\,\ref{fig:1}\,(c) in the plane which includes $\mathbf{\Gamma^{\prime\prime}}$ and $\mathbf{R^{\prime\prime}}$ points and which is perpendicular to the $k_y$-$k_z$ plane. In the latter case the distribution of the Berry curvature field, although similar to the previous case, is more complex, owing to the fact that the crossings at $\mathbf{\Gamma^{\prime}}$ and $\mathbf{R^{\prime}}$ are very close to the plane so that the overall distribution coming from all four points is plotted.

 \begin{figure}[b]
    \centering\vspace{-1pt}
    \includegraphics[width=\linewidth]{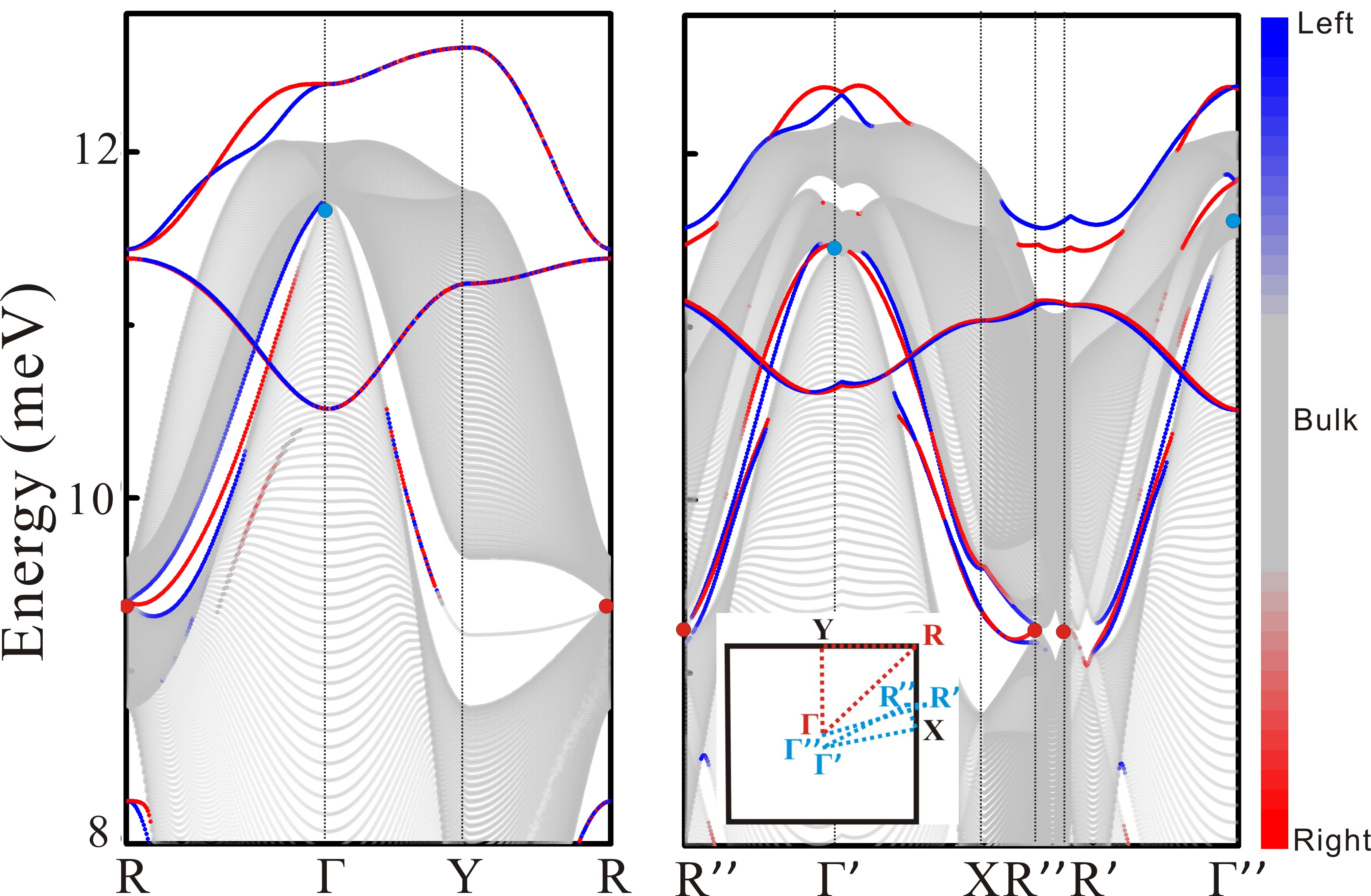}
    \caption{The surface magnon band structure of the 75-layer thick slab of Cu$_2$OSeO$_3$(001)
    along the paths indicated in the inset shown on the left  without the DMI, and on the right including the DMI. The special points $\mathbf{\Gamma}$, $\mathbf{R}$, $\mathbf{\Gamma^{\prime}}$, $\mathbf{R^{\prime}}$, $\mathbf{\Gamma^{\prime\prime}}$, and $\mathbf{R^{\prime\prime}}$ are the projections of the Weyl points onto the $(001)$ plane, which are indicated with red (positive charge) and blue (negative charge)  solid circles. The color scale represents the weight of the magnonic wave function along the slab.\vspace{-3pt}}
    \label{fig:3}
\end{figure}

Next, we compute the monopole charge of the $i$'th band crossing by evaluating the flux of the cumulative Berry curvature field through an infinitesimal two-dimensional sphere $S_i$ surrounding the crossing:
\begin{equation}
 Q_i=\frac{1}{2\pi}\int_{S_i}\boldsymbol{\mathrm{\Omega}}(\mathbf{k})\cdot\mathbf n \,dS_i,
\end{equation}
where $\mathbf{n}$ is the surface normal. According to our calculations, without the DMI, the total topological charge of the two degenerate points at $\mathbf{\Gamma}$ is $+2$,  while it constitutes a value of $-2$ at $\mathbf{R}$. Upon including the effect of the DMI, each of the double degeneracies splits into two nondegenerate points with charges of $+1$ at $\mathbf{\Gamma^{\prime}}$ and $\mathbf{\Gamma^{\prime\prime}}$, and $-1$ at $\mathbf{R^{\prime}}$ and $\mathbf{R^{\prime\prime}}$. The topological analysis is further supported by the Brillouin zone evolution of the first Chern number, defined analogously to the charge as:
\begin{equation}
 C(P)=\frac{1}{2\pi}\int_{P}\boldsymbol{\mathrm{\Omega}}(\mathbf{k})\cdot\mathbf n \,dP,
\end{equation}
where $P$ is a two-dimensional slice of the Brillouin zone and $\mathbf{n}$ is its normal. By defining the plane $P$ as the $k_x$-$k_y$ plane at a given $k_z$ with $\mathbf{n}=(0,0,1)$, we compute the evolution of $C(k_z)$ as a function of $k_z$, presenting the results in Fig.~S5 of the Supplemental Material \cite{SupplementalMaterial}. Without DMI, the Chern number changes by 2 when $P$ passes through the degenerate crossing points, while in the presence of DMI it changes by 1 when $P$ passes through every nondegenerate crossing point. This analysis underlines the main finding of our paper\,---\,the emergence of two doubly degenerate type-I Weyl points\,\cite{SoluyanovGresch15} in the magnonic structure of Cu$_2$OSeO$_3$, located at $\mathbf{R}$ and $\mathbf{\Gamma}$ without DMI, which further split into overall four Weyl points when the symmetry of the system is reduced by including the DMI into consideration.


\vspace{-2pt}\subsection{\hspace{-1ex}Surface states}\vspace{-2pt}

As the emergence of the Weyl points in the magnonic band structure of a three-dimensional crystal is expected to give rise to the surface states of a thin film, here, we analyze the magnon band structure of a 75-layer thick two-dimensional slab of Cu$_2$OSeO$_3$ cut along the [001] axis, presenting the results in Fig.\,\ref{fig:3}.
The spin-wave dispersion is shown along the path which includes the projections of the Weyl points onto $(001)$-plane, which are further indicated with red and blue small circles in the figure, according to their topological charge. In the magnon band structure, the states are marked with their weight at the surface of the slab (see Supplemental Material \cite{SupplementalMaterial} for more details). The left plot in Fig.\,\ref{fig:3} corresponds to the situation without the DMI, with projections of the Weyl points positioned at high symmetry points in the two-dimensional Brillouin zone. We observe that in this case the Weyl points of opposite chirality are connected by the magnon ``arc'' surface states, which is in accord with our topological analysis from above. Upon including the effect of the DMI, the Weyl points split, and their projections move to the $\mathbf{\Gamma^{\prime}}$, $\mathbf{R^{\prime}}$, $\mathbf{\Gamma^{\prime\prime}}$ and $\mathbf{R^{\prime\prime}}$ points. Again, this is consistent with the previous analysis of the topological charges: While the points of the same charge are not connected by the surface states, the points of opposite chirality are. Additional analysis of the surface magnon arcs and surface band structure is given in the Supplemental Material \cite{SupplementalMaterial}.



\subsection{\hspace{-1ex}Thermal Hall conductivity}\vspace{-2pt}

The topological thermal Hall effect of magnons is the generation of a transverse thermal Hall voltage under an applied longitudinal temperature gradient due to the presence of the DMI~\cite{OnoseIdeue10, HirschbergerChisnell15}.
The energy-dependent contribution to the $i{\kern-.5pt}j$'th Cartesian component of the thermal Hall conductivity tensor $\hat{\kappa}$ can be calculated as
\begin{equation}
 \kappa^{ij}(\epsilon)=-\frac{k_\text{B}^2 T}{(2\pi)^3 \hbar} \sum_n\int_\text{BZ}\delta(\epsilon_{n\mathbf{k}}-\epsilon)\,C_2 (f_n^\text{B})\, \Omega^{ij}_n(\mathbf{k})\,d\mathbf{k},
 \label{equ2}
\end{equation}
where $n$ enumerates the magnon bands, $f_n^\text{B}$ is the Bose-Einstein distribution function, which can be expressed as $f_n^\text{B}=(e^{\epsilon_{n\mathbf{k}}/k_\text{B}T}-1)^{-1}$, and $C_2$ is given by
\begin{equation}
C_2(x)=(1+x)\left(\ln\frac{1+x}{x}\right)^2-\ln^2 x-2\mathrm{Li}_2 (-x),
\end{equation}
with  $\rm{\ Li}_2$ denoting the dilogarithm function. The thermal Hall conductivity tensor of the system is then defined as
$ \kappa^{ij}=\lim_{\mu\rightarrow\infty}\kappa^{ij}_{\mu}$, where $\kappa^{ij}_{\mu}=\int_0^{\mu}\kappa^{ij}(\epsilon)\,d\varepsilon$ is the cumulative thermal Hall conductivity.

\begin{figure}[!t]
    \centering
    \includegraphics[width=\linewidth]{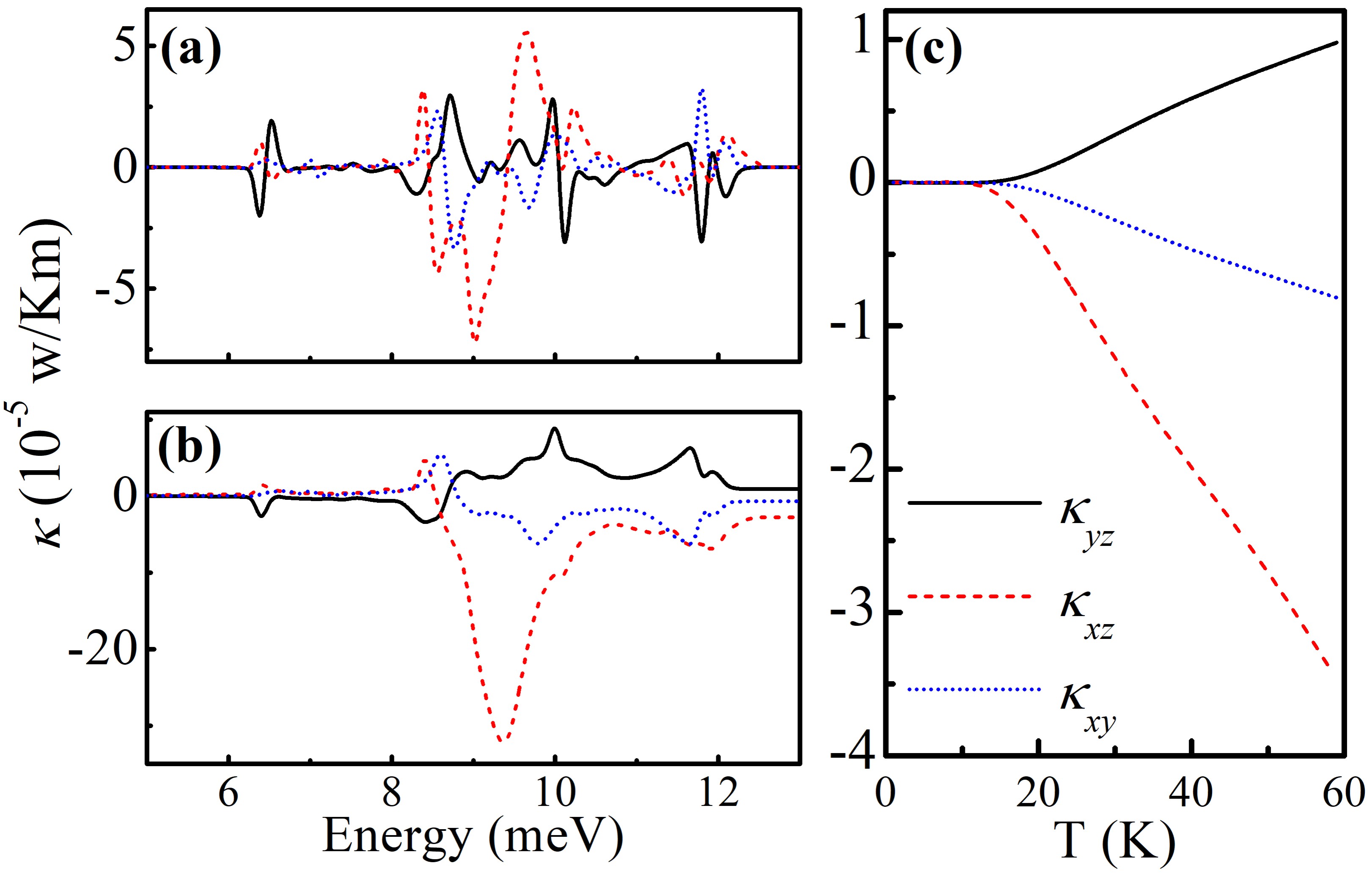}
    \caption{Components of the thermal Hall conductivity tensor in Cu$_2$OSeO$_3$. Energy-dependent, (a), and cumulative, (b), thermal Hall conductivity computed at 60\,K. (c) The temperature dependence of the thermal Hall conductivity. }
    \label{fig:4}
\end{figure}
From experiment we know that the Curie temperature of Cu$_2$OSeO$_3$ is around $60\,\mathrm{K}$~\cite{Janson14, TuckerWhite16}. The computed energy-dependence and the cumulative components of the thermal Hall conductivity, calculated according to equations above at 60\,K, are shown in Figs.\,\ref{fig:4}\,(a,b). In these plots we observe that in the energy region  between 9 and 10\,meV there is a significant enhancement especially in the $\kappa^{xz}$ component of the thermal Hall conductivity. This enhancement can be attributed to the distribution of the Berry curvature around the Weyl points in that energy region, which correspondingly gives rise to the fingerprint of the Weyl points in the energy distribution of the thermal Hall effect. Since the Weyl-point enhancement is most prominent for the $\kappa^{xz}$ component, the overall value of the thermal Hall conductivity for this component is by far dominant over other two components at 60\,K, see Figs.\,\ref{fig:4}\,(c), where the
thermal Hall conductivity as a function of temperature is shown. As  magnons obey the Bose-Einstein distribution, and the low-lying states are thus responsible for the thermal Hall effect at low temperatures, the characteristic zero-plateau in $\kappa$ observed in Figs.\,\ref{fig:4}\,(c) is the consequence of the vanishing contribution by the ``topologically-trivial'' low-lying bands which are basically not affected by the DMI,
Fig.~\ref{fig:1}(d). Respectively, the thermal Hall effect ``lifts off'' once the region of Weyl points is reached by the distribution of magnons. The overall magnitude of the thermal Hall effect that we predict in the region of higher temperatures is large enough to be observed in experiment.


\subsection{\hspace{-1ex}Effect of the DMI on the position of Weyl points}\vspace{-2pt}
\begin{figure}[!t]
    \centering
    \includegraphics[width=\linewidth]{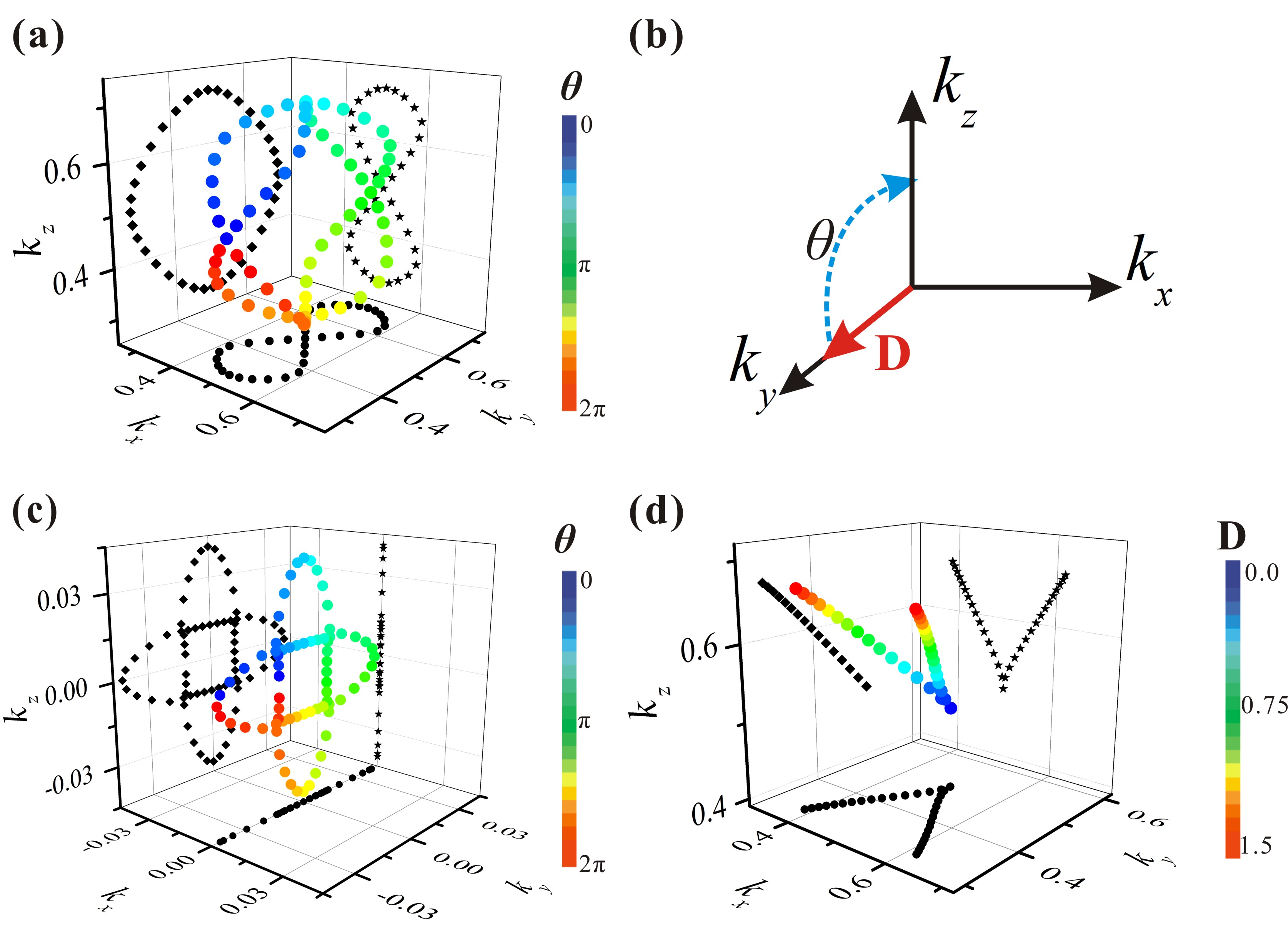}
    \caption{The effect of the nearest neighbor Dzyaloshinsky-Moriya coupling on the positions of the Weyl points. The DMI vector is rotated about the $x$ axis with the initial direction  along [010], (b), and the evolution of the corresponding Weyl points with angle $\theta$ near $\mathbf{R}$ (a) and $\mathbf{\Gamma}$ (c) points is shown. The projection of each Weyl point onto the $k_x$-$k_y$, $k_x$-$k_z$ and $k_y$-$k_z$ planes is shown with black symbols in the corresponding planes. The evolution of the Weyl points near $\mathbf{R}$ as a function of the DMI strength, with the DMI vector along [110], is shown in (d). The color scale in (d) corresponds to the magnitude of the DMI in meV.}
    \label{fig:5}
\end{figure}
Given the low structural symmetry of Cu$_2$OSeO$_3$, it seems reasonable to explore the influence of the direction and strength of the DMI vector on the position of the Weyl points in the Brillouin zone. While we envisage that the tuning of the DMI parameters can be realized~e.g.~by pressure, strain, electric field~\cite{wells2017effect,nii2015uniaxial,srivastava2018large,koyama2018electric},  or doping with defects, knowing the correlation between the Weyl point geometry and the DMI provides a unique tool for accessing the details of the DMI in a given sample, which are challenging to extract with other techniques based e.g. on measuring the properties of domain walls~\cite{ryu2013chiral,emori2013current}.

To estimate the influence of the DMI on the Weyl points, we first keep the direction of the DMI along the [110] axis, while scaling its magnitude between 0 and 1.5\,meV (color scale in Fig.\,\ref{fig:5}\,(d)).
The evolution of the Weyl points around
$\mathbf{R}$ upon increasing the DMI is shown in Fig.\,\ref{fig:5}\,(d). Notably, upon starting from a degenerate case at zero DMI, the splitting between the two Weyl point is clearly driven by lowering of symmetry upon including the non-vanishing DMI, while the trajectories of the Weyl points in the Brillouin zone are almost perfectly straight lines.

 Further, after fixing the magnitude of the DMI to the value of 1meV, we rotate the direction of the nearest neighbor DMI vector, as specified by angle $\theta$ in Fig.\,\ref{fig:5}(b), about the $x$-axis, and track the position of two Weyl points around $\mathbf{\Gamma}$ and $\mathbf{R}$ in Fig.\,\ref{fig:5}(c) and (a), respectively. The results indicate that the Weyl points rotate around the $\mathbf{R}$ and $\mathbf{\Gamma}$ points along specific paths when following the rotation of the DMI vector. The corresponding trajectories, while having a relatively complex shape in the three-dimensional Brillouin zone, clearly possess a high degree of symmetry, as apparent from the projections of the trajectories onto the high-symmetry planes, see e.g. Fig.\,\ref{fig:5}(a) and (c).
 We show further data on the correlation between the DMI and the Weyl point behavior in the Supplemental Material \cite{SupplementalMaterial}.


\vspace{-5pt}\section{\hspace{-1ex}Discussion}\label{Sec:Discussion}

In our work, based on the spin-wave theory and experiment, we arrived at several important findings concerning the spin-wave properties of ferrimagnetic Cu$_2$OSeO$_3$. Firstly, we were able to attribute the origin of the experimentally observed magnon band gap in the spin-wave spectrum at the $\mathbf{R}$ point to the effect of the DMI, which was chosen so as to provide the best fit to the high-resolution neutron scattering data.  Secondly, after systematically addressing the topological properties of Cu$_2$OSeO$_3$, we uncovered the emergence of the doubly-degenerate Weyl nodes with topological charge $\pm2$ at high-symmetry points even without the effect of the DMI. We further observed that each Weyl point splits into two as the symmetry of the system is reduced when introducing the DMI into play. Importantly, we find that the position of the Weyl points can be controlled by changing the crystal symmetry of the compound. We further predict that the emergence of the Weyl points in the system goes hand in hand with the formation of topological magnonic surface states, which can be observed for instance at the (001) surface of Cu$_2$OSeO$_3$.

Our findings open a quest for experimental observation of the Weyl points in this material, and exploring the influence of such points in the spin-wave spectrum on various properties of more complex magnetic phases in Cu$_2$OSeO$_3$, for example, its skyrmion phase. While we discover that Weyl points play a crucial role in shaping the magnitude and temperature dependence of the thermal Hall effect in its ferrimagnetic phase, we expect that the same holds true also for skyrmions in Cu$_2$OSeO$_3$. The observation of the exact position of the Weyl points as well as following their dynamics upon structural reconstructions in Cu$_2$OSeO$_3$ can further provide a unique tool for accessing the microscopics of the DMI in this complex compound, which can be of paramount importance for understanding and shaping of chiral dynamics and properties of Cu$_2$OSeO$_3$. The latter finding also suggests that in special materials of Cu$_2$OSeO$_3$ type one can expect that the topologies in the space of magnons and in the real-space of skyrmions can be closely intertwined.\smallskip\smallskip

\section{Acknowledgements}\label{Sec:Acknowledgements}
We acknowledge fruitful discussions with Flaviano Jos\'{e} dos Santos, Sergii Grytsiuk, Matthias Redies, Lizhi Zhang, and Wanxiang Feng. This project was supported by the China Scholarship Council (CSC) (Grant No.~[2016]3100) and by the German Research Foundation (DFG) within the Collaborative Research Center SFB\,1143 at the TU Dresden (project C03), the individual research grants IN 209/4-1, MO 1731/5-1, and the priority programme SPP~2137 ``Skyrmionics'' (projects IN~209/7-1 and MO 1731/7-1). This work has been also supported by the DFG through the Collaborative Research Center SFB 1238. We gratefully acknowledge computing time on the supercomputers of J\"ulich Supercomputing Center, and at the JARA-HPC cluster of RWTH Aachen.

\bibliographystyle{my-apsrev}
\bibliography{Cu2OSeO3}\vspace{-2pt}

\onecolumngrid

\clearpage
\setcounter{section}{0}
\setcounter{figure}{0}

\makeatletter\renewcommand{\fnum@figure}[1]{\figurename~S\thefigure.}\makeatother
\makeatletter\renewcommand{\fnum@table}[1]{\tablename~S\thetable.}\makeatother

\begin{center}
\noindent{\large \textbf{Supplemental Material for the Article\\ ``Magnonic Weyl states in Cu$_2$OSeO$_3$''}}
\end{center}

\begin{center}
\noindent L.~Zhang, Y.~A. Onykiienko, P.~M.~Buhl, Y.~V. Tymoshenko, P. \v{C}erm\'{a}k,\\ A.~Schneidewind, A.~Henschel, M.~Schmidt, S.~Bl\"ugel, D.~S.~Inosov, Y.~Mokrousov
\end{center}\bigskip

\twocolumngrid

\section{DMI vector}
\begin{figure}[htbp]
    \centering
    \includegraphics[width=7.5cm]{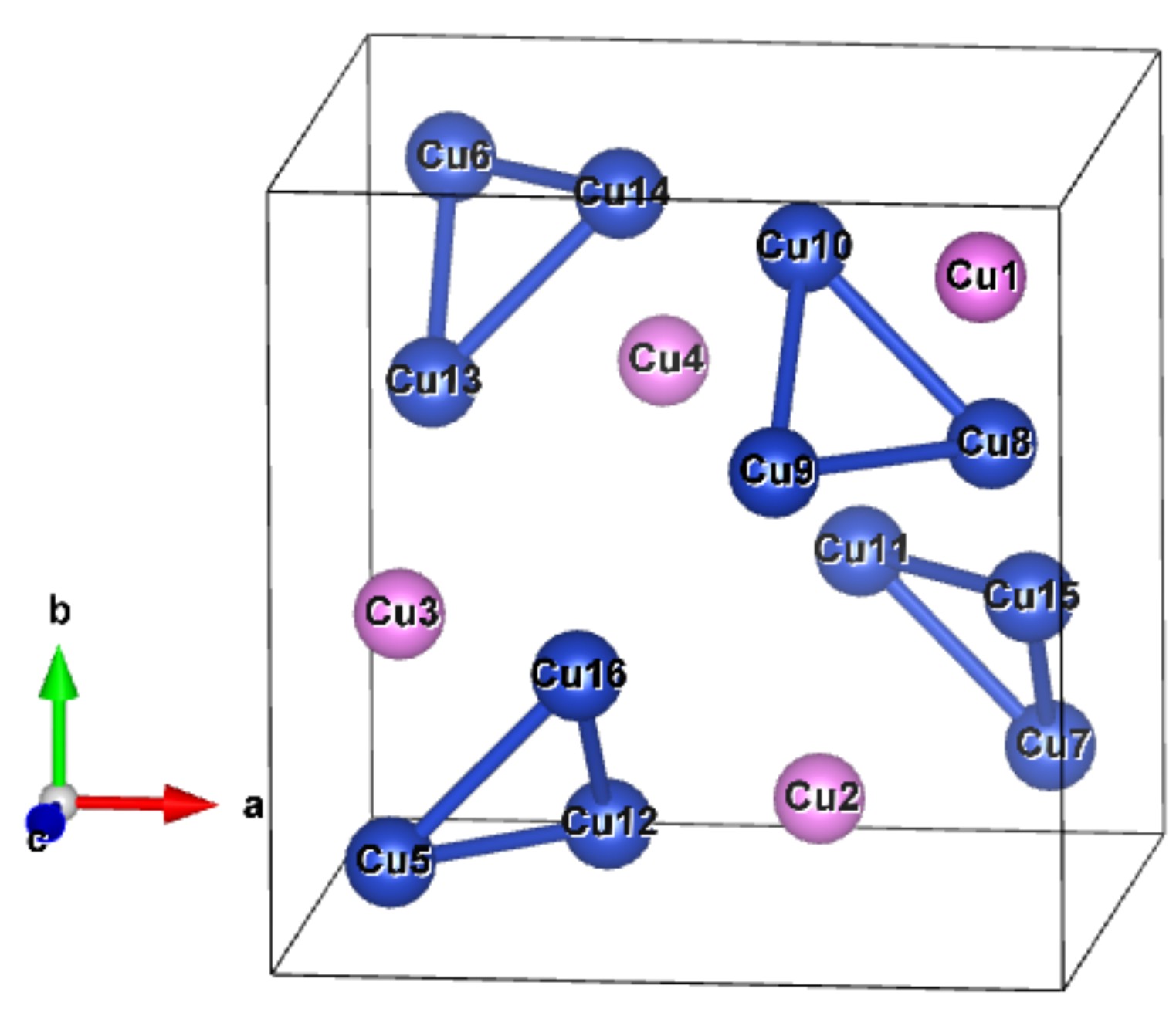}\vspace{-3pt}
    \caption{The atomic positions of copper atoms in the unit cell of Cu$_2$OSeO$_3$. The blue lines represent the nearest neighbor bonds of Table~S\ref{table:1}.\vspace{-3pt}}
    \label{figs:1}
\end{figure}

\vspace{-2pt}
The cubic Cu$_2$OSeO$_3$ has the B$20$ structure with the space group $P2_{1}3$.
Employing the symmetry operation, the positions of all Cu$^{2+}$ ions can be obtained from the positions of the Cu$(1)$ $(u,u,u)$ and the Cu$(2)$ $(a,b,c)$ atom, where $u=0.88557$, $a=0.13479$, $b=0.12096$, and $c=0.87267$.
Furthermore, the global symmetries relate the different bond directions of equivalent pairwise interactions. The corresponding DMI vectors obey the same respective symmetry relations as the bond directions.
The atomic positions, the bond directions, and the corresponding DMI vectors of the nearest neighbour interaction are listed in  Table~S\ref{table:1}, the copper structure and the nearest neighbor interaction bonds are additionally visualized in Fig.\,S\ref{figs:1}.
The table is restricted to the nearest neighbor DMI since only this interaction is considered in the main text with $\mathbf{D}=(-0.458, 2.011, 0.565)$~meV. Further DMI vector  relations of different neighbours can be obtained through similar symmetry operations.

\begin{table*}[htbp]
\caption{The bond vector and the DMI vector for the nearest neighbor interaction.}
\begin{tabular}{l@{\qquad}c@{\qquad}c@{\qquad}c}
 \toprule
Atom $i$ position & Atom $j$ position & $\mathbf{R}_{ij}$  &  $\mathbf{D}_{ij}$\\
\midrule
$r_5$, $(a,b,c)$ & $r_{12}$, $(0.5-b, 1-c,0.5+a)$	& $(0.5-b-a, 1-c-b, 0.5+a-c)$	& $(D_x, D_y, D_z)$\\
$r_5$, $(a,b,c)$& $r_{16}$, $(-0.5+c, 0.5-a,1-b)$ & $(-0.5+c-a,0.5-a-b,1-b-c)$	& $(-D_z, D_x, D_y)$\\
$r_6$, $(b,c,a)$& $r_{13}$, $(1-c, 0.5+a,0.5-b)$&	$(1-c-b,0.5+a-c,0.5-b-a)$&	$(D_y, D_z, D_x)$\\
$r_6$, $(b,c,a)$ & $r_{14}$, $(0.5-a, 1-b, -0.5+c)$& $(0.5-a-b,1-b-c, -0.5+c-a)$&	$(D_x, D_y, -D_z)$\\
$r_7$, $(c,a,b)$ &	$r_{11}$, $(0.5+a, 0.5-b,1-c)$&	$(0.5+a-c,0.5-b-a,1-c-b)$&	$(D_z, D_x, D_y)$\\
$r_7$, $(c,a,b)$ &	$r_{15}$, $(1-b, -0.5+c,0.5-a)$& $(1-b-c, -0.5+c-a,0.5-a-b)$&	$(D_y, D_z, D_x)$\\
$r_8$, $(1-a,0.5+b, 1.5-c)$ &	$r_9$, $(0.5+b, 1.5-c,1-a)$&	$(-0.5+b+a,1-c-b, -0.5-a+c)$	&$(-D_x, D_y, -D_z)$\\
$r_8$, $(1-a,0.5+b, 1.5-c)$ &	$r_{10}$, $(1.5-c, 1-a,0.5+b)$&	$(0.5-c+a,0.5-a-b, -1+b+c)$&	$(D_z, D_x, -D_y)$\\
$r_9$, $(0.5+b, 1.5-c,1-a)$ &	$r_{10}$, $(1.5-c, 1-a,0.5+b)$&	$(1-c-b, -0.5-a+c, -0.5+b+a)$&	$(D_y, -D_z, -D_x)$\\
$r_{11}$, $(0.5+a, 0.5-b,1-c)$ &	$r_{15}$, $(1-b, -0.5+c,0.5-a)$&	$(0.5-b-a, -1+c+b,-0.5-a+c)$&	$(D_x, -D_y, -D_z)$\\
$r_{12}$, $(0.5-b, 1-c,0.5+a)$ &	$r_{16}$, $(-0.5+c, 0.5-a,1-b)$&	$(-1+c+b, -0.5-a+c,0.5-b-a)$&	$(-D_y, -D_z, D_x)$\\
$r_{13}$, $(1-c, 0.5+a,0.5-b)$ & $r_{14}$, $(0.5-a, 1-b, -0.5+c)$&	$(-0.5-a+c,0.5-b-a, -1+c+b)$&	$(-D_z, D_x, -D_y)$\\
\bottomrule\vspace{-3pt}
\label{table:1}
\end{tabular}
\end{table*}

\begin{figure}[t]
    \centering
    \includegraphics[width=\columnwidth]{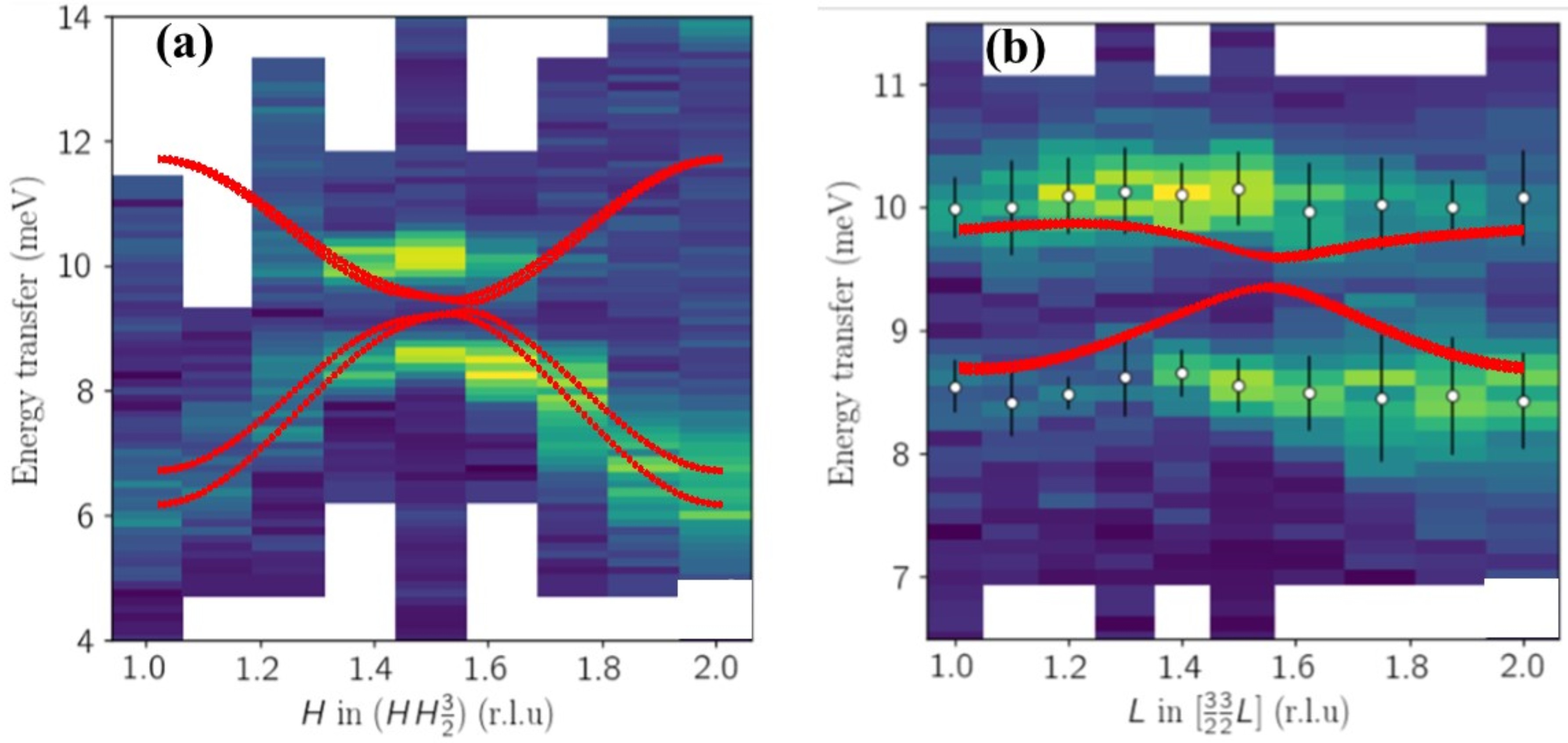}
    \caption{Momentum energy cuts along (a)~(110) and (b)~(001) directions that cross at (1.5,1.5,1.5). The measurements were done at 2~K without applied magnetic field. The red dotted lines represent the theoretical results using the referenced, \textit{ab-initio} DMI vectors.}
    \label{figs:2}
\end{figure}

\begin{figure}[b]
    \centering
    \includegraphics[width=7CM]{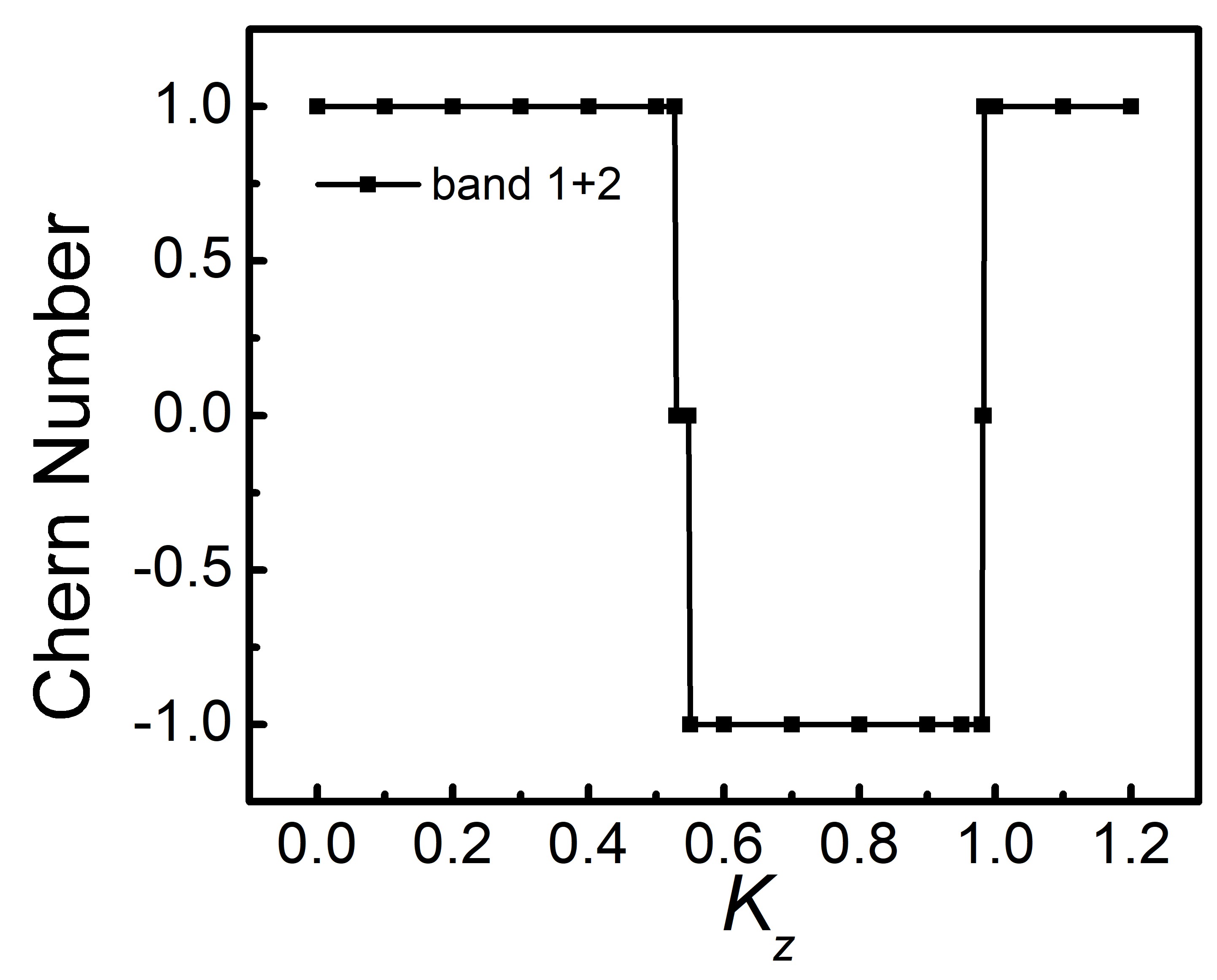}
    \caption{The integer Chern number sum of bands 1 and 2 as a function of $k_z$ computed with {\it ab-initio} values of the DMI.\vspace{-3pt}}
    \label{figs:3}
\end{figure}

\section{Results with \textit{ab-initio} parameters}\vspace{-2pt}
In this paper DMI are introduced to enhance the agreement between the experimental magnon dispersion and the LSWT calculations. While the nearest neighbour DMI of the main text was obtained through fitting to the measured dispersion, this section utilizes $5$ distinct \textit{ab-initio} DMI which were reported by O.~Janson \textit{et al.} [Nat. Commun. \textbf{5}, 5376 (2014)].
This study also supplied the Heisenberg interaction parameters used in the main text with the spin up moment 0.450 and spin down moment 0.483, which is very similar to the \textit{ab-initio} results.
The resulting magnon dispersion on top of the experimental data is shown in Fig.~S2.

Using the  {\it ab-initio} DMI parameters, four Dirac points were found in the first Brillouin zone (BZ) at the following positions: (0.568, 0.464, 0.549), (0.568, 0.507, 0.528), (0.014, 0.020, 0.983), and (0.020, 0.022, $-$0.019). Their topological characters are evaluated similar to the calculations corresponding to Eq.\,7 of the main text. Accordingly, the Chern numbers of two-dimensional (2D) Brillouin zone slices perpendicular to the (001)-direction are plotted as a function of $k_z$ in Fig.\,S\ref{figs:3}.
The quantized changes of the Chern number at the Dirac point positions reveal the same topological characteristics as the Weyl points of the main text. Hence, this system exhibits Weyl points even when including further DMI.\vspace{-1pt}

\section{Fitting to experimental spectrum}\vspace{-2pt}

\begin{figure}[b]
    \centering
    \includegraphics[width=8.5cm]{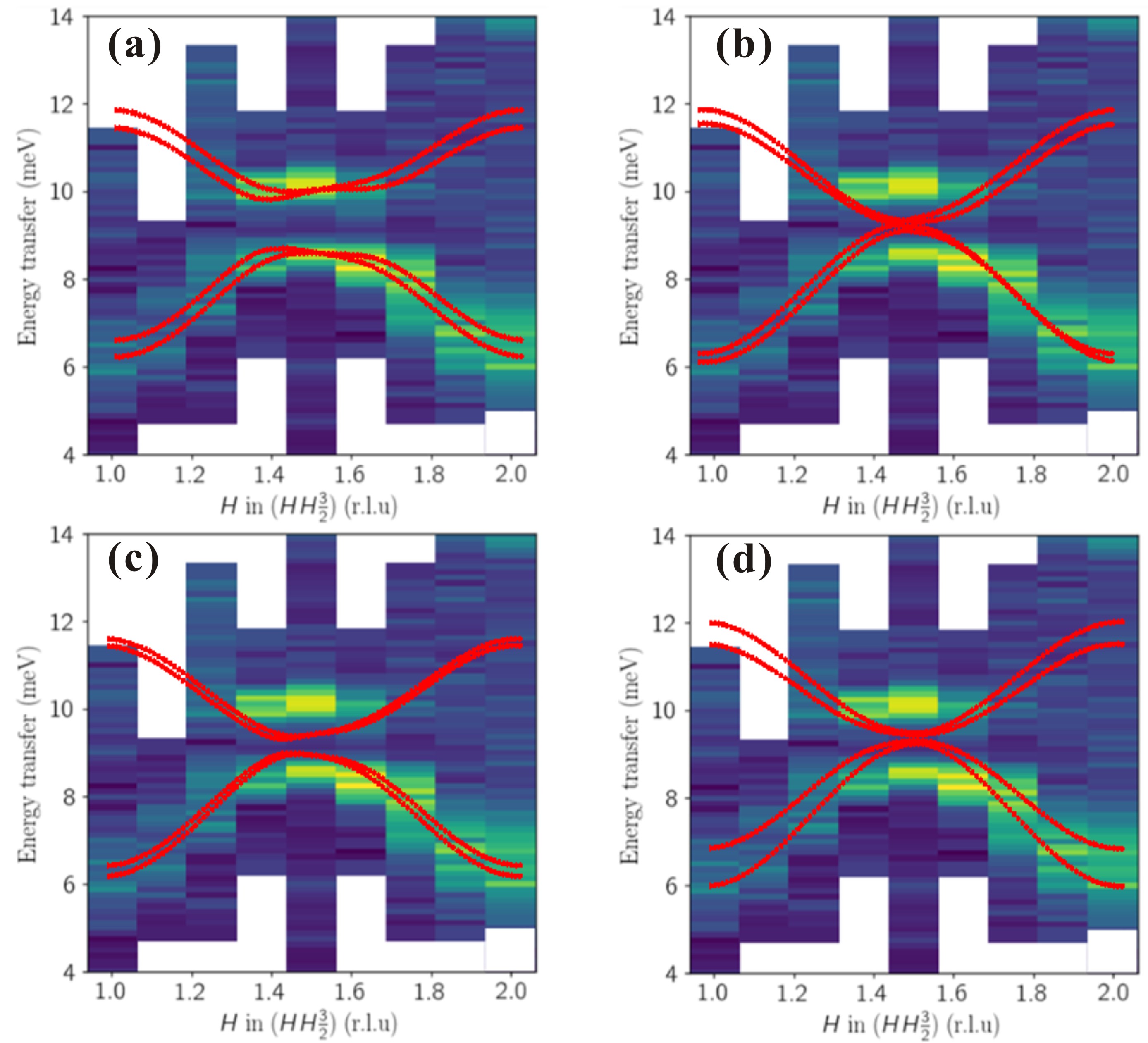}
    \caption{The comparison of fitted results to the experimental data. Panels (a)--(d) correspond to the results obtained when considering only  the first, second, third, and fourth nearest neighbor DMI vector, respectively.}
    \label{figs:4}
\end{figure}
 Alternatively, the DMI vectors can be obtained by fitting to the experimental magnon dispersion. This section describes the procedure as applied in the main text and demonstrates analogous results of further interactions.
 First, $13$ representative points of experimental data set were selected to comprise the fitting measure. Subsequently, the Broyden-Fletcher-Goldfarb-Shanno (BFGS) Hessian update strategy is employed to converge the considered DMI vector.
 This procedure is executed separately for each DMI vector of the first $4$ nearest neighbors given in Table~1 of the main text.
 After few iteration steps, the first, second, third and fourth nearest DMI vectors converge to
$(-0.458, 2.011, 0.565)$ meV, $(1.505, 6.143, 3.013)$ meV,
$(0.262, 4.615, 1.236)$ meV and $(3.976, -2.484, -2.024)$ meV, respectively. Comparison of the corresponding magnon dispersions with the experimental results (Fig.\,S\ref{figs:4}) reveals great accuracy when considering the first nearest neighbor DMI but uncovers poor agreement in all other cases. This justifies the restriction to the first DMI in the main text.


\section{Chern number evolution}

\begin{figure}[h]
    \centering\vspace{2pt}
    \includegraphics[width=7.4cm]{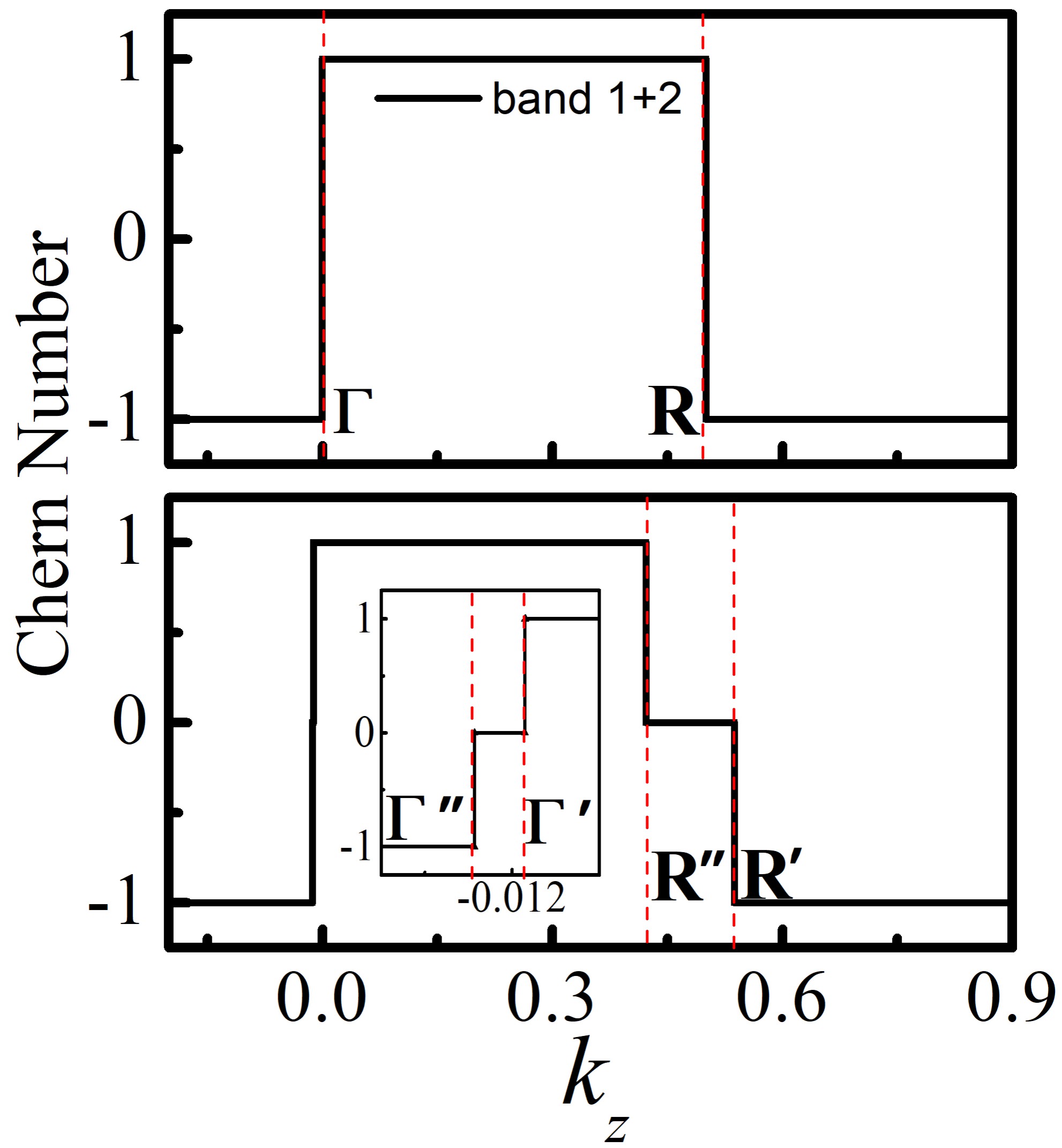}\vspace{-3pt}
    \caption{The evolution of the Chern number as a function of $k_z$. The upper and lower panels display the result without and with DMI effect, respectively. The inset is the zoom into the region around  $k_z=-0.012$.\vspace{-6pt}}
    \label{figs:5}
\end{figure}

The Chern number was calculated in order to analyze the topological character of the Weyl points. Based on Eq.\,7 of the main text, the Chern number sum of bands 1 and 2 was computed for Brillouin zone slices perpendicular to $k_z$. The upper and lower plots of Fig.\,S\ref{figs:5} display the Chern number dependence on $k_z$ (for details see main text) without and with DMI, respectively. The Weyl point positions are marked by red dashed lines.

\section{Surface localization and the surface arc}

The color scale in Fig.\,$3$ of the main text is calculated based on the following equation:
 \begin{equation}
 \label{eq:local}
 LW(k,j)=\sum_iV_L^i(k,j)V_R^i(k,j)(R_z^i-0.5),
\end{equation}
where $k$ is the reciprocal space vector, $j$ denotes the band index,  $i$ numbers the magnetic atom, and $R^i_z$ represents the normalized position for atom $i$ along the $z$-axis. $V_L^i(k,j)$ and $V_R^i(k,j)$ are the components of the left and right eigenstates of $j$ at the magnetic atom $i$.

\begin{figure}[t]
    \centering
    \includegraphics[width=\columnwidth]{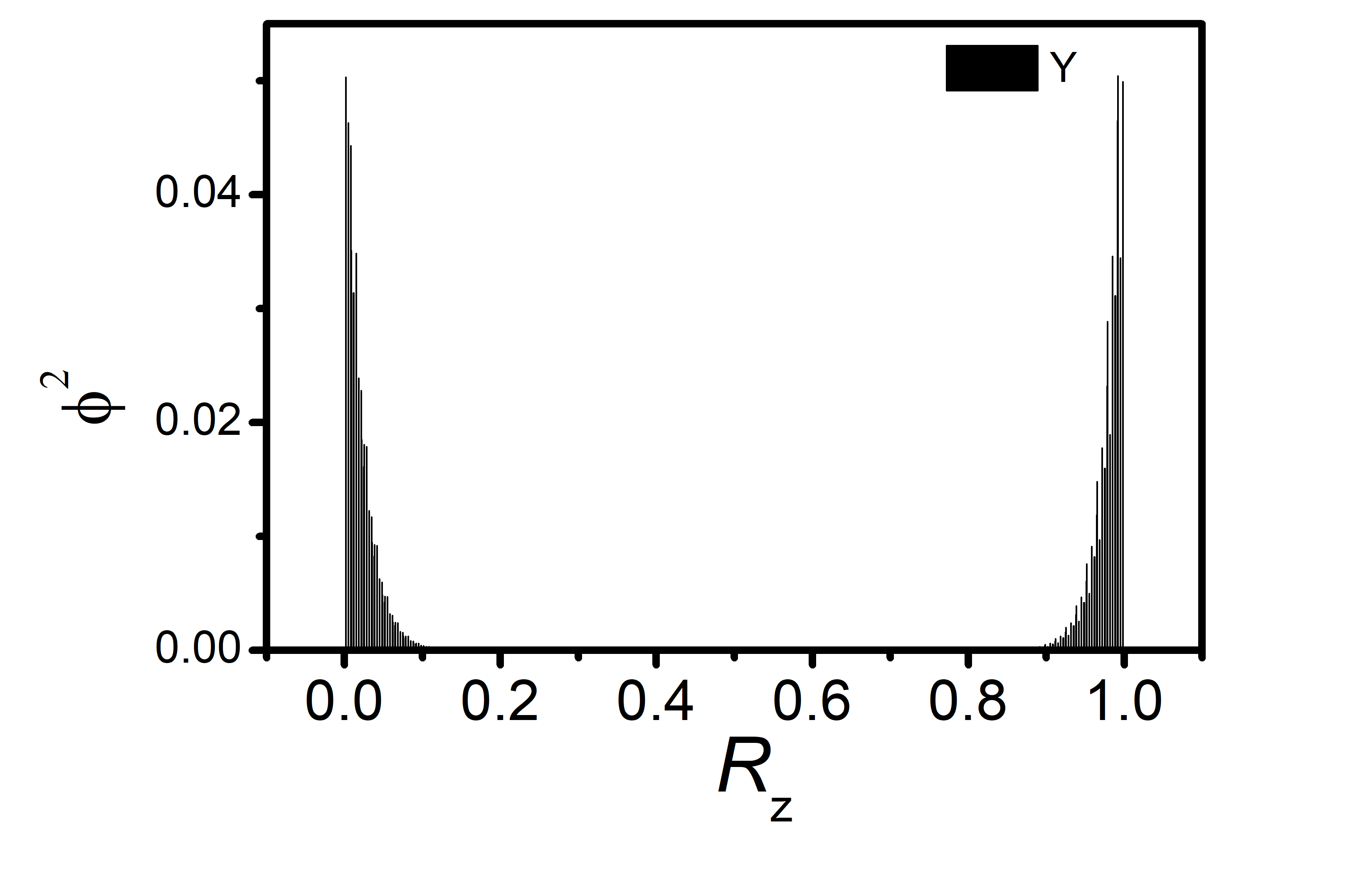}\vspace{-6pt}
    \caption{This figure visualizes the real-space distribution of the weight of the $Y$-point magnon at $9.2$\,meV of Fig.\,3 (left) of the main text. DMI is not included and 75 layers (1200 atoms) were used in the calculation.\vspace{-3pt}}
    \label{figs:6}
\end{figure}

The above expression provides a reasonable measure to judge the surface character of each state while distinguishing both sides. However, in the left plot of Fig.\,3 (main text) surface states at about 9.2~meV along $\overline{\Gamma Y}$ and $\overline{Y R}$ appear to loose and regain their surface character without contact to bulk states, which is highly unusual. This issue is resolved by Fig.\,\ref{figs:6} which shows the real-space decomposition of one of these apparent bulk states at the $Y$ point. That state exhibits highly localized contributions on both surfaces but none in the bulk, hence the degenerate states are indeed highly localized at the surfaces. Accordingly, Eq.\,\ref{eq:local} is unable to classify such states equally localized on both surfaces. This justifies the assumption that the other apparent bulk states along that high symmetry line are of surface character as well. Numerically, the origin of the surface character concealment is the exact energetic degeneracy which causes unsuitable eigenstate superpositions.



\begin{figure}[h]
    \includegraphics[width=7cm]{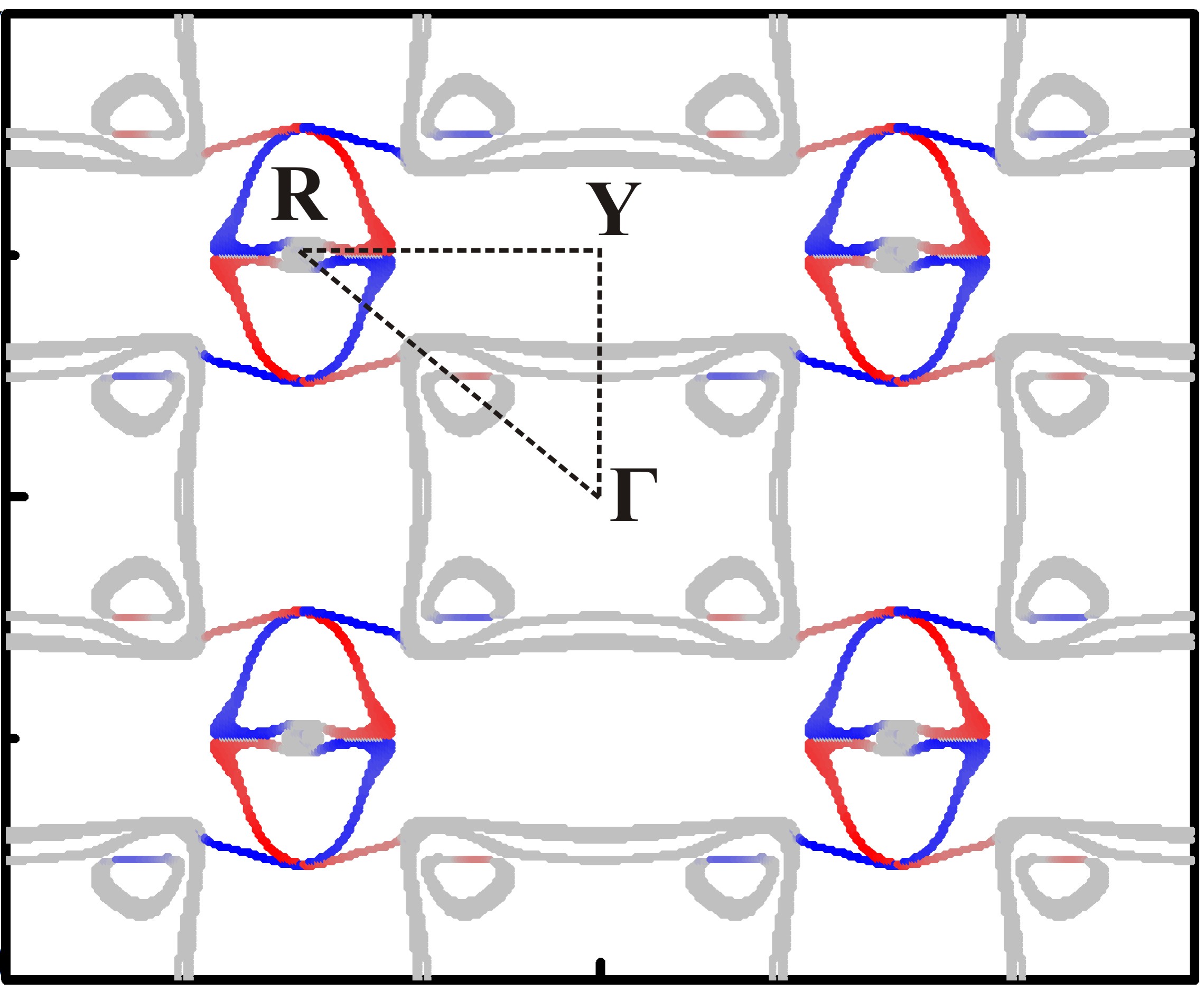}
    \caption{The surface arcs calculated without DMI effect for the selected energy of 9.25~meV.}
    \label{figs:7}
\end{figure}

Fig.\,S\ref{figs:7} shows the distribution of the surface states in the surface Brillouin zone at the energy of 9.25~meV. The blue and red represent the states at the left and right surfaces of a 20-layer thick slab, while the grey color indicates bulk states. 
The plot visualizes clear arcs connecting the projected Weyl point to bulk states which enclose the Weyl points of opposite character.

\section{DMI effect on the positions of the Weyl points}\vspace{-2pt}

\begin{figure*}[t!]
    \centering\vspace{5pt}
    \includegraphics[width=\textwidth]{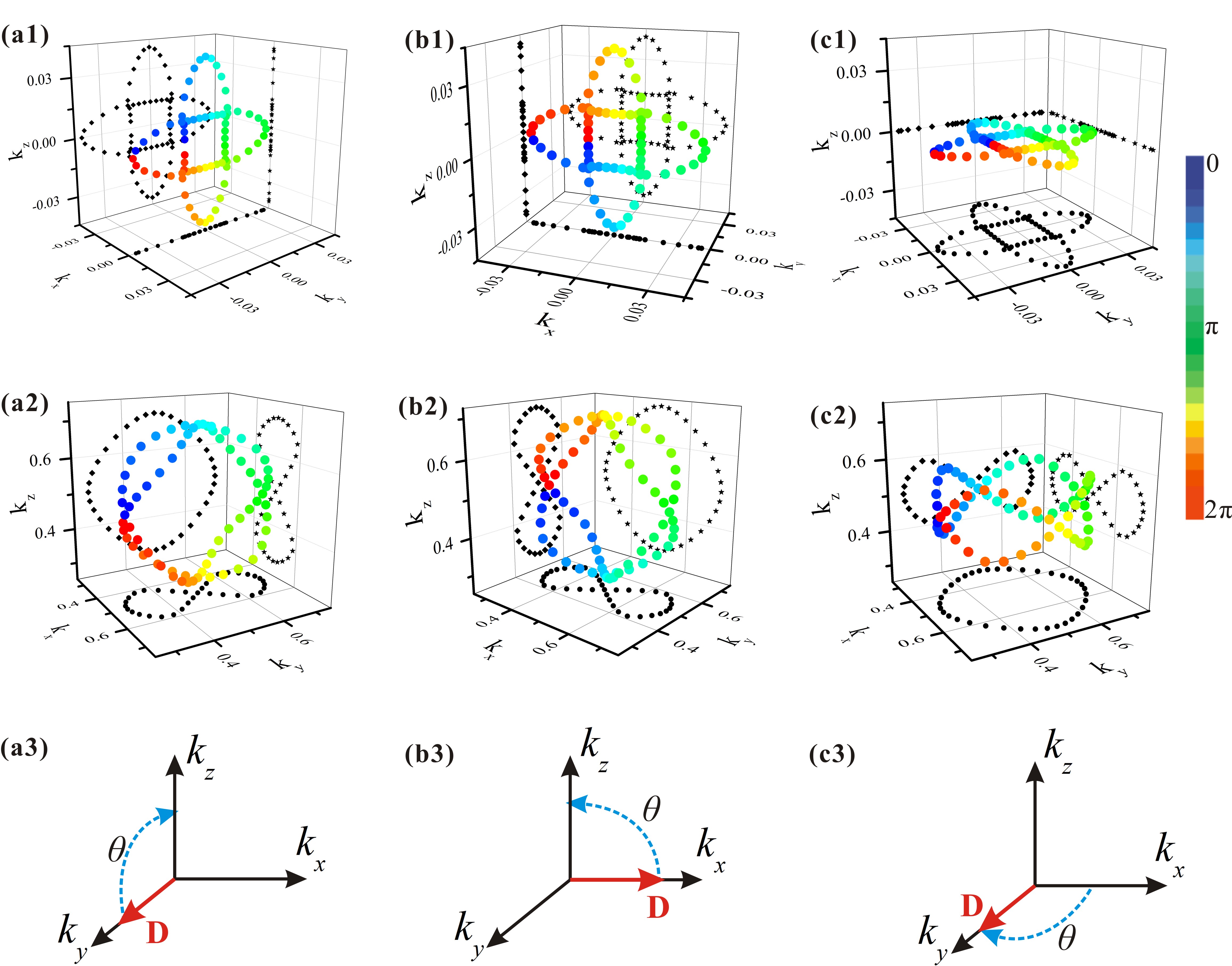}
    \caption{The effect of the  DMI  on  Weyl point positions. Only the nearest DMI was taken into account and the positions of Weyl points were drawn separately near the $\mathbf{R}$ (a2, b2, c2) point and $\mathbf{\Gamma}$ (a1, b1, c1) points. (a,b,c) corresponds to the rotation of the  DMI vector around $x$, $y$ and $z$ axes with the initial vector along $[010]$ (a3), $[100]$ (b3)  and $[010]$ (c3).
    The color map represents the value of $\theta$ in the range from $0$ to 2$\pi$.\vspace{-5pt}}
    \label{figs:8}
\end{figure*}

Fig.\,S\ref{figs:8} on the next page demonstrates further correlations between the nearest neighbor DMI and the Weyl point positions supplementing Fig.\,5 of the main text and the corresponding section. Here, the DMI vector is rotated around the $x$, $y$, and $z$ axes starting with initial orientation along the $[010]$-, $[100]$- and $[010]$-direction, respectively. Highly symmetric Weyl point trajectories are uncovered demonstrating at least approximate symmetry between the unit directions on this level.

\onecolumngrid

\end{document}